\documentclass[aps,prl,twocolumn,superscriptaddress,reprint]{revtex4-1}

\usepackage[ansinew]{inputenc}
\usepackage[T1]{fontenc}
\usepackage{ae,aecompl}
\usepackage[english]{babel}
\usepackage{graphicx,pict2e}
\usepackage{hyperref,amsmath}
\usepackage{epstopdf}
\usepackage{color}
\usepackage{multirow}

\begin{document}

\author{F. Arnold}
\author{M. Naumann}
\author{S.-C. Wu}
\author{Y. Sun}
\author{M. Schmidt}
\author{H. Borrmann}
\author{C. Felser}
\affiliation{Max Planck Institute for Chemical Physics of Solids, 01187 Dresden, Germany} 
\author{B. Yan}
\affiliation{Max Planck Institute for Chemical Physics of Solids, 01187 Dresden, Germany}
\affiliation{Max Planck Institute for Physics of Complex Systems, 01187 Dresden, Germany}
\author{E. Hassinger}
\affiliation{Max Planck Institute for Chemical Physics of Solids, 01187 Dresden, Germany}

\title{Chiral Quasiparticles at the Fermi Surface of the Weyl Semimetal TaAs}

\date{today}

\begin{abstract}

Tantalum arsenide is a member of the non-centrosymmetric monopnictides, which are putative Weyl semimetals. In these materials, three-dimensional chiral massless quasiparticles, the so-called Weyl fermions, are predicted to induce novel quantum mechanical phenomena, such as the chiral anomaly and topological surface states. However, their chirality is only well-defined if the Fermi level is close enough to the Weyl points that separate Fermi surface pockets of opposite chirality exist. In this article, we present the bulk Fermi surface topology of high quality single crystals of TaAs, as determined by angle-dependent Shubnikov-de Haas and de Haas-van Alphen measurements combined with \textit{ab-initio} band-structure calculations. Quantum oscillations originating from three different types of Fermi surface pocket were found in magnetization, magnetic torque, and magnetoresistance measurements performed in magnetic fields up to $14\,\mathrm{T}$ and temperatures down to $1.8\,\mathrm{K}$. Of these Fermi pockets, two are pairs of topologically non-trivial electron pockets around the Weyl points and one is a trivial hole pocket. Unlike the other members of the non-centrosymmetric monopnictides, TaAs is the first Weyl semimetal candidate with the Fermi energy sufficiently close to both types of Weyl points to generate chiral quasiparticles at the Fermi surface.

\end{abstract}

\maketitle

The non-centrosymmetric monophosphides (TaP, NbP) and monoarsenides (TaAs, NbAs) are members of a new class of topological materials, the Weyl semimetals. These are thought to host new exotic quantum mechanical phenomena such as topological surface states \cite{Wan11} and the chiral transport anomaly \cite{Alder69,Bell69} due to the presence of three-dimensional chiral massless Weyl fermions \cite{Weyl29, Volovik12}. Quasiparticles in semimetals behave as Weyl fermions if the bands disperse linearly in reciprocal space near three dimensional band crossing points (Weyl nodes), and the spin is locked to the wave vector $\boldsymbol k$ such that the chirality is given by the sign of $\boldsymbol \sigma \cdot \boldsymbol k$, where $\boldsymbol \sigma_i$ are the Pauli matrices. This can be realized in semimetals with strong spin-orbit coupling and broken inversion or time-reversal symmetry, which induces pairs of Weyl nodes of opposite chirality \cite{Weng15,Huang15}. 

Probes sensitive to the Fermi surface (FS), such as specific heat, magnetization and electrical transport are expected to show unconventional behavior related to massless chiral fermions only if the Fermi energy $E_\mathrm{F}$ is close to the Weyl points. More precisely, this window of energies is defined by the existence of separate Fermi pockets around each Weyl point (Weyl pockets), giving rise to quasiparticles of well-defined chirality. Detuning the Fermi energy beyond this range leads to the merging of neighboring Weyl pockets and cancellation of the associated Berry flux and chirality. Thus precise knowledge of the band structure and Fermi energy are needed in order to attribute observed anomalous quantum phenomena to chiral Weyl fermions. 

\begin{figure}[!b]
	\centering
		\includegraphics[width=0.95\columnwidth]{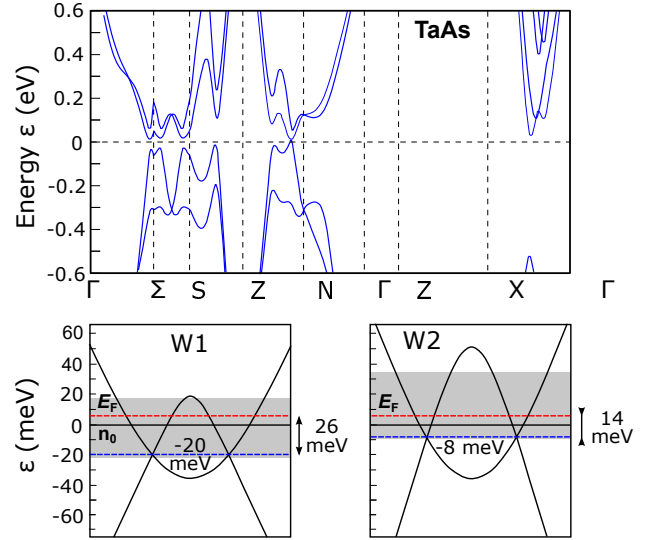}
			\centering
	\caption{Top: \textit{Ab-initio} band structure of TaAs. Bottom: Scheme of the bands close to the Fermi energy along the line connecting the Weyl nodes. $n_0$ marks the charge neutral point, $E_\mathrm{F}$ the Fermi energy as determined by quantum oscillations. The grey-shaded area corresponds to the Fermi energy window in which separate Fermi surface pockets of opposite chirality exist.} 
	\label{fig:Bandstructure}
\end{figure}

\begin{figure*}[!t]
\centering
\includegraphics[width=0.95\textwidth]{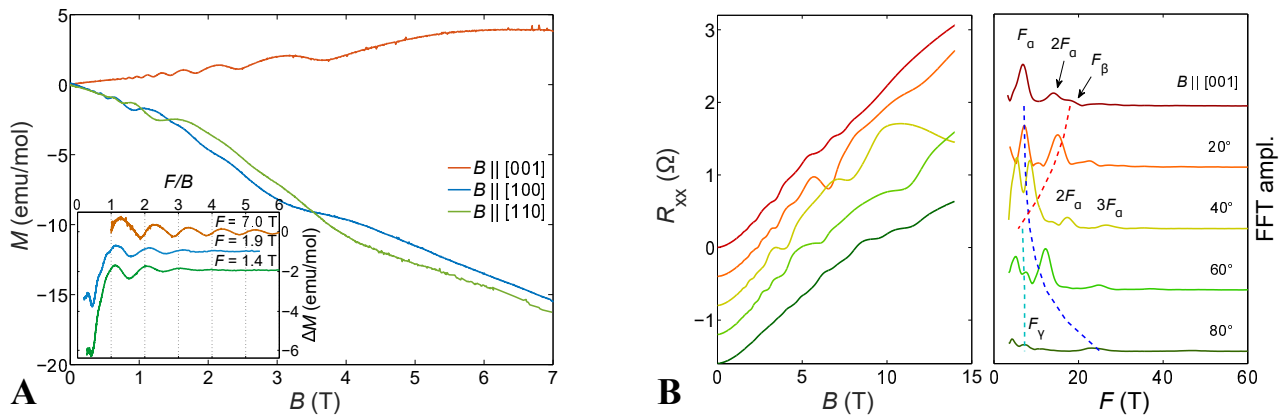}
\caption{A) Magnetic field dependence of the magnetization at $2\,\mathrm{K}$. The inset shows the de Haas-van Alphen oscillations plotted over the inverse magnetic field normalized by the main quantum oscillation frequency. A linear background fitted to the low field magnetization was subtracted. Curves are offset for clarity. B) left panel: transverse magnetoresistance  for magnetic fields applied within the (100)-plane in steps of $20^o$ measured at $2\,\mathrm{K}$ (data have been offset by $0.4\,\Omega$ each for clarity). right panel: corresponding Fourier transformations of the background subtracted Shubnikov-de Haas signals over a magnetic field range of 1 to $14\,\mathrm{T}$. The dashed lines are guides to the eye showing the angular dependence of $F_\alpha$, $F_\beta$ and $F_\gamma$.}
\label{fig:Data}
\end{figure*}

Early angle-resolved photo emission spectroscopy (ARPES) studies observed Fermi arcs on the sample surface of TaP and TaAs and linearly dispersing bands in the bulk, proving the existence of two sets of Weyl nodes (W1 and W2) in these materials \cite{Lv15,Xu15,Yang15NatPhys}. However, due to the limited energy resolution of ARPES, the precise Fermi energy location with respect to the Weyl points and the resulting FS topology remained an open question. Recently Klotz \textit{et al.} and Arnold \textit{et al.} reported on the bulk FS topology of NbP and TaP, respectively \cite{Klotz16,Arnold15TaP}. It was shown that the Fermi energy in both materials is well outside the range for observing chiral Weyl fermions, and the FS consists only of topologically trivial pockets. 

Here we present the first demonstration of Fermi-level Weyl fermions of opposite chirality in a monoarsenide or monophosphide, by showing that in as-grown crystals of TaAs $E_\mathrm{F}$ is within the window of separate FS pockets around each Weyk point. A full angular dependence study of the de Haas-van Alphen (dHvA) and Shubnikov-de Haas (SdH) effect of high quality single crystals of TaAs was performed by means of magnetization, magnetic torque and electrical transport measurements. Unlike previous studies \cite{Zhang15,Huang15PRX}, we obtain a unique solution for the FS topology by considering hitherto undiscovered quantum oscillation frequencies of the minor Weyl and hole pockets, which allow us to locate the Fermi energy with millielectronvolt precision. 

\begin{figure*}[!t]
\centering
\includegraphics[width=0.98\textwidth]{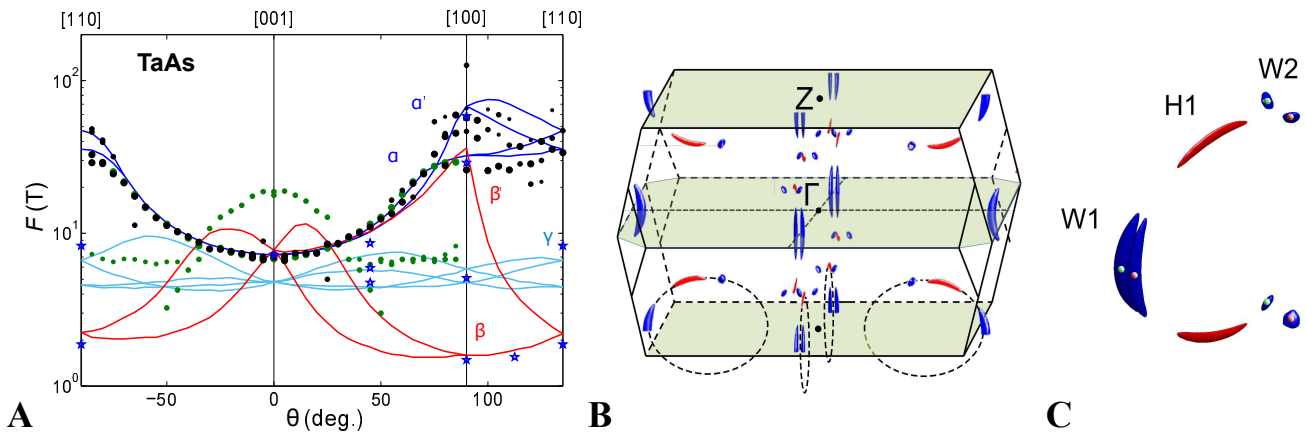}
\caption{A) Angular dependence of the quantum oscillation frequencies $F$ of TaAs for magnetic fields applied within the (100), (110) and (001)-plane. Green and black dots as well as blue stars represent frequencies determined from electrical transport, magnetic torque and magnetization respectively. The solid lines show the theoretical angular dependence according to the \textit{ab-initio} band structure at a Fermi energy of $+6\,\mathrm{eV}$. B and C) First Brillouin zone and Fermi surface topology of TaAs according to the angular dependence in graph A). The dashed rings show the nodal lines as explained in the text.}
	\label{fig:AngularDependence}
\end{figure*}

The band structure of TaAs was calculated using the Vienna \textit{ab-initio} simulation  package (VASP) \cite{Kresse96} (see Fig. \ref{fig:Bandstructure}). The modified Becke-Johnson exchange potential \cite{Becke06,Tran09} was employed for accurate band structure calculations. FSs were interpolated using maximally localized Wannier functions \cite{Mostofi08} in dense k-grids (equivalent to $300\times 300\times300$ points over the entire BZ). In the absence of spin-orbit coupling, the valence and conduction band cross each other along four so called 'nodal rings'. These rings lie in the (100) and (010) mirror planes (see dashed rings in Fig. \ref{fig:AngularDependence}B) \cite{Weng15}. SOC is found to gap out the nodal ring except for the band crossings at the Weyl points. The grey-shaded bands in Fig. \ref{fig:Bandstructure} give the energy range where separate Weyl pockets can be found in TaAs.

Crystals of TaAs were grown by chemical vapor transport starting from micro-crystalline powder. The initial TaAs powder was synthesized by reacting tantalum (Chempur $99.9\,\%$) and arsenic (Alfa Aesar $99.9999\,\%$) in an evacuated fused silica tube for 24 hours at $600\,^o\mathrm{C}$ and 24 hours at $800\,^o\mathrm{C}$ \cite{Martin88}. Subsequently, large single crystals were grown by vapor transport in a temperature gradient between $1000\,^o\mathrm{C}$ (source) an $900\,^o\mathrm{C}$  (sink) using iodine (Alfa Aesar $99,998\,\%$) as transport agent with a concentration of $13\,\mathrm{mg/cm}^3$. The resulting crystals showed a residual resistivity ratio of 10 (see Fig. S3 of the Supplemental Online Material (SOM) \cite{SOM}). Their crystalline structure, lattice constants and alignment were confirmed using x-ray diffraction techniques \cite{SOM}.

Quantum oscillations (QOs), the dHvA and SdH effect, arise due to the $1/B$-periodic depopulation of Landau levels (LLs) at $E_\mathrm{F}$ with increasing magnetic field \cite{Shoenberg}. Their frequencies are related to the extremal FS cross section $F= \hbar/2\pi e A_\mathrm{extr}$ normal to the magnetic field, and can be used to reconstruct the FS topology of metals\cite{Onsager52}. 

The magnetization of TaAs along all major axes and selected angles in between were measured in a QD $7\,\mathrm{T}$ SQUID-VSM at temperatures between 2 and $20\,\mathrm{K}$ (see Fig. \ref{fig:Data}A and Fig. S6 of the SOM \cite{SOM}).
Angle-dependent SdH and dHvA measurements were performed in a $14\,\mathrm{T}$ Quantum Design (QD) Physical-Property Measurement System (PPMS) using the QD rotator probe and TqMag piezoelectric torque cantilevers \cite{Rossel96}. Measurements were taken at $1.8\,\mathrm{K}$ while the magnetic field was swept from $+14\,\mathrm{T}$ to $-14\,\mathrm{T}$ (SdH) and $+14\,\mathrm{T}$ to zero field (dHvA) at a rate of $80\,\mathrm{Oe/s}$ at highest and $10\,\mathrm{Oe/s}$ at lowest fields. Resistance and off-balance signals were measured using external Stanford Research SR830 and SR850 lock-in amplifiers. Typical excitation currents of 1 to $5\,\mathrm{mA}$ at 16 to $22\,\mathrm{Hz}$ were applied to the sample and TqMag cantilever. Raw resistance and torque data for magnetic fields applied within the (100) and (110)-plane can be found in Fig. S5 of the SOM \cite{SOM}.

Magnetization data and their background subtracted dHvA oscillations for $B\|[100]$, $[110]$ and $[001]$ at $2\,\mathrm{K}$ are shown in Fig. \ref{fig:Data}A. As can be seen, TaAs shows a diamagnetic magnetization along the $[100]$ and $[110]$-axis, and paramagnetic magnetization along the $[001]$-axis. Strong QOs are visible in all three directions. This is in contrast to NbP and TaP where the magnetization is diamagnetic for all field directions and quantum oscillations are weak for magnetic fields applied in the $(001)$-plane \cite{Klotz16,Arnold15TaP}. The inset of Fig. \ref{fig:Data}A shows the background subtracted dHvA oscillations plotted against $F/B$. The observed low frequencies are indicative of very small FS pockets as are typical for semimetals with low charge carrier densities. The frequency of $7.0\,\mathrm{T}$ for $B\|[001]$ agrees well with previous reports by other groups \cite{Zhang15,Huang15PRX,Zhang16}. In addition, we find a step-like magnetization feature along the $[100]$ and $[110]$-direction starting at $F/B=1$ or equivalently 1.4 and $1.9\,\mathrm{T}$ which is attributed to the quantum limit of the lowest frequency. At this step the dominant frequency vanishes and a second larger frequency of another FS pocket becomes visible. 
 
The detailed angular dependence of the QO frequencies was determined by combined electrical transport (see e.g. Fig. \ref{fig:Data}B) and magnetic torque measurements. Fourier transforms of multiple field windows and measurement techniques were correlated to produce the angular dependence shown in Fig. \ref{fig:AngularDependence}A. Table \ref{tab:Summary} contains a summary of all QO frequencies along the major axes. 

By tuning $E_\mathrm{F}$ to $+6\,\mathrm{meV}$ in our MBJ band structure (Fig. \ref{fig:Bandstructure}), we can fit our measured QO frequencies with the predicted angular dependence of the extremal orbit size (Fig. \ref{fig:AngularDependence}A). This energy is well within the energy window for observing chiral FS pockets (see Fig. \ref{fig:Bandstructure}) and the FS is described by three Weyl pocket pairs, one W1 and two W2 pairs. These coexist with two trivial hole pockets, H1, aligned along the nodal ring (see Fig. \ref{fig:AngularDependence}B and C). Crucially, a unique solution of the FS topology was only achieved by including all three QO frequency branches. This way it was possible to eliminate alternative topologies deep in the hole and electron doped regime, which give rise to a similar $1/\cos(\theta)$ dependence of the dominant $7\,\mathrm{T}$ frequency (see Fig. S10 of the SOM \cite{SOM}). The tetragonal crystal symmetry and consequential $90^\circ$ repetition of the FS pockets around the [001]-axis gives rise to a frequency splitting when the magnetic field is tilted away from the [001]-axis.

\begin{figure}[!tbp]
	\centering
		\includegraphics[width=0.9\columnwidth]{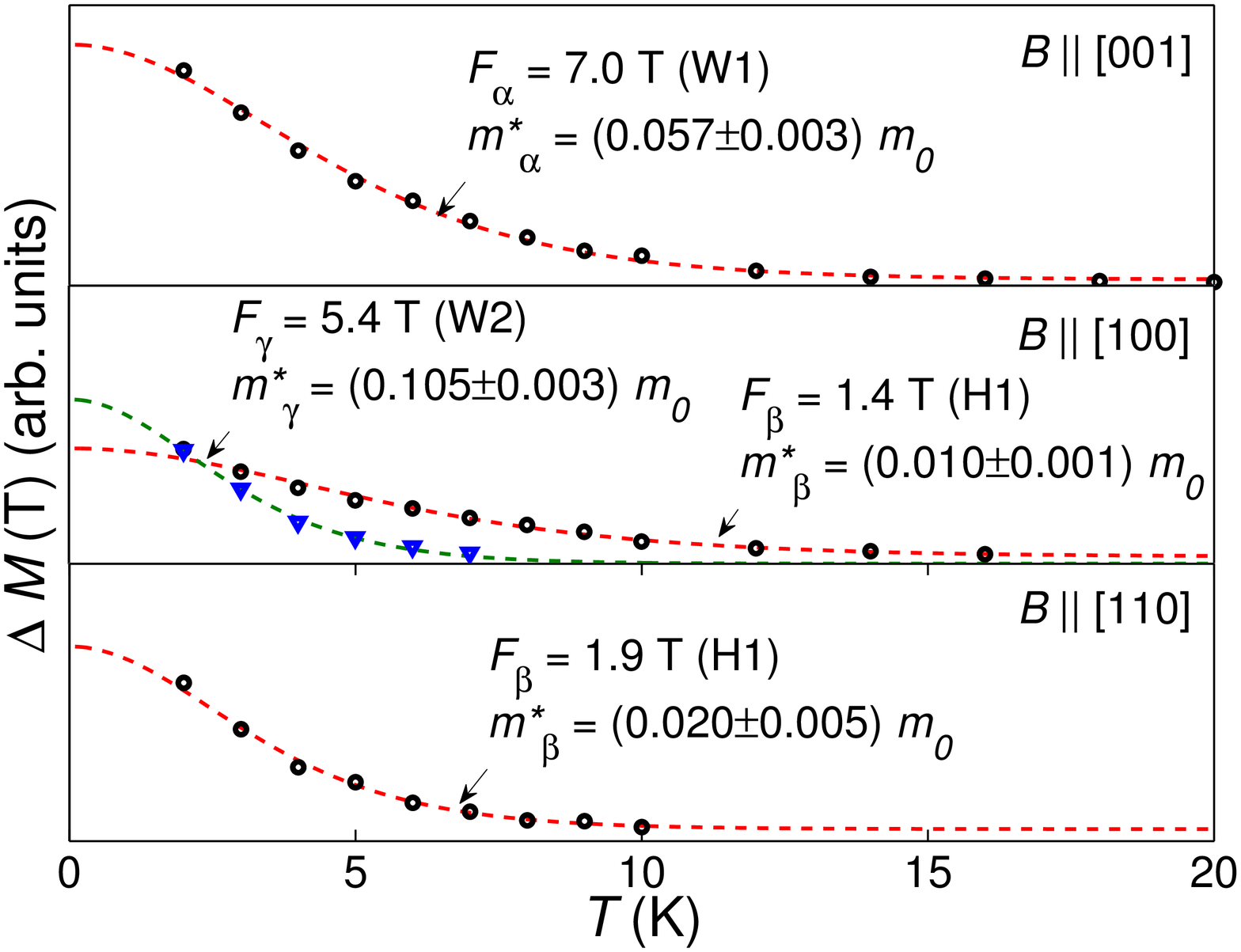}
	\caption{Temperature dependences of the de Haas-van Alphen oscillation amplitude as determined from magnetization measurements in the magnetic field interval of 1 to $7\,\mathrm{T}$ along the high symmetry axes. The dashed lines are fits to the Lifshitz-Kosevich temperature reduction term $\tilde{M}\propto x/\sinh{(x)}$ with $x=\pi^2m^*k_BT/\mu_BB$ \cite{Lifshitz56,Lifshitz58,Shoenberg}.}
	\label{fig:LifshitzKosevichFits}
\end{figure}

The banana-shaped electron W1 pockets at the zone boundary, which are aligned parallel to the $[001]$-axis, give rise to an inverse $\cos(\theta)$-like angular dependence of the $F_\alpha$ branch (dark blue lines in Fig. \ref{fig:AngularDependence}A) when the magnetic field is tilted away from the [001]-axis. The extremal cross section of W1 is equivalent to $7\,\mathrm{T}$ for $B\|[001]$ and increases up to 29 and $60\,\mathrm{T}$ for $B\|[110]$. The corresponding aspect ratio of these pockets is approximately $1:7$ (mean $(001)$-plane diameter:$[001]$-axis length). The charge carrier effective mass of the according electrons has been determined by fits of the Lifshitz-Kosevich temperature reduction term (see Fig. \ref{fig:LifshitzKosevichFits}) and yields $(0.057\pm0.003)\,m_0$ for $B\|[001]$ where $m_0$ is the free electron mass \cite{Lifshitz56,Lifshitz58}. This is in excellent agreement with the effective mass of 0.06 predicted by our band structure calculation \cite{SOM}.

The W2 pockets on the other hand are rather isotropic spheres. Their angular dependence is split into four branches due to their off-planar position and low symmetry. The corresponding frequency $F_\gamma$ (light blue lines in Fig. \ref{fig:AngularDependence}A) is almost independent of the magnetic field direction. The effective mass for $B\|[100]$ has been determined from the visible magnetization oscillations between $1.2$ and $3.0\,\mathrm{T}$ (see Fig. \ref{fig:Data}A) as $(0.105\pm0.003)\,\mathrm{m}_0$. 

The H1 hole pockets give rise to the $F_\beta$ branch (red lines in Figure \ref{fig:AngularDependence}A). $F_\beta$ starts at $19\,\mathrm{T}$ for $B\|[001]$ and drops quickly to $F_\beta=1.4\,\mathrm{T}$ and $1.9\,\mathrm{T}$ along the $[100]$ and $[110]$-axis. This is indicative of a strongly anisotropic ellipsoidal pocket aligned almost parallel to the $[100]$-axes. Its aspect ratio is approximately 1:10. Experimentally, we observe a continuous decrease of the extremal orbit size when the magnetic field is tilted away from the $[001]$-axis. The calculations on the other hand shows an initial increase, peaking around $12^\circ$, towards the a-axis. This initial increase is due to the non-horizontal alignment of the hole pocket in the Brillouin zone. The observed discrepancy can originate from a stronger bend or fine structure of the hole pocket, which is beyond the accuracy of our calculations and stabilizes a larger orbit for magnetic fields along the c-axis.The fitted effective masses of the H1 pocket are $(0.010\pm0.001)\,m_0$ and $(0.020\pm 0.005)\,m_0$ for $B\|[100]$ and $B\|[110]$ respectively (see Fig. \ref{fig:LifshitzKosevichFits}). 
As we saw in the magnetization data, it is the H1 pocket which reaches its quantum limit above $2\,\mathrm{T}$ for magnetic fields within the $(001)$-plane. The magnetization steps at $F_\beta/B=1$ along the $[100]$ and $[110]$-axis correspond to the second last LL crossing $E_\mathrm{F}$. Beyond this field only the $n=0$ LL remains occupied. The quantum limit field scales with the QO frequency and $F_\beta[110]\approx\sqrt{2}F_\beta[100]$, as expected for a quasi-two dimensional FS aligned parallel to the $[100]$-axis. Thus no quantum limit signature is expected in the magnetic field range up to $7\,\mathrm{T}$ for $B\|c$. As the electrical transport is dominated by the larger W1 and W2 pockets, which are still described by multiple LLs in this field range, no clear transport anomalies are observed. 


\begin{table}[tb]
\centering
\caption{Experimental and calculated effective cyclotron masses $m^*$ and quantum oscillation frequencies $F$ for different extremal orbits and magnetic field orientations. $m_0$ is the free electron mass.}
\begin{tabular}{c c c c c c c} 
\hline
\hline
Pocket & \multicolumn{2}{c}{Orbit}  & \multicolumn{2}{c}{Experiment} & \multicolumn{2}{c}{Calculation}\\
 & & & $F$ (T) & $m^*$ $(m_0)$& $F$ (T) & $m^*$ $(m_0)$\\
\hline
W1 & $\alpha$ & $B \| [001]$ & 7.0 & 0.057(3)  &7.25 & 0.065\\
\multirow{2}{*}{W2} & $\gamma$ & $B \| [100]$ & 5.4(3) & 0.105(3) & 5.3(5) &0.076(6)\\
 & $\gamma$ & $B \| [110]$ & 8.2(1) & - & 5.5(11) & 0.082(14)\\
\multirow{3}{*}{H1} & $\beta$ &$B \| [001]$ &18.75 &- & 7.77 & 0.17\\
 & $\beta$ &$B \| [100]$ & 1.4(1) & 0.010(1) & 1.59 & 0.024\\
 & $\beta$ &$B \| [110]$ & 1.9(1) & 0.020(5) &2.20 & 0.035\\
\hline
\hline
\end{tabular}
\label{tab:Summary}
\end{table}

As the measured $E_\mathrm{F}$ lies $6\,\mathrm{meV}$ above the charge neutral point (see Fig. \ref{fig:Bandstructure}) the system is slightly electron doped. By comparing the hole and electron FS volumina we obtain a theoretical electron doping of $3.5\times10^{18}\,\mathrm{cm}^{-3}$. This electron doping is also seen as a negative Hall resistance (Fig. S4 of the SOM \cite{SOM}). A single band analysis of the high field Hall resistance yields an electron doping of $(1.0\pm0.5)\times10^{19}\,\mathrm{cm}^{-3}$, which is in reasonable agreement with our QO measurements. Contrary to an ideal compensated semimetal the observed doping indicates slight deviations from a defect free stoichiometric crystal. In general, different growth conditions do not seem to affect the QO frequencies of TaAs and thus its doping level and $E_\mathrm{F}$ \cite{Zhang15,Huang15PRX,Zhang16}. 

A Dingle analysis of the W1 and H1 dHvA oscillations (see Fig. S7 of the SOM \cite{SOM}) reveals scattering times of $\tau(W1)=3.8\times10^{-13}\,\mathrm{s}$ and $\tau(H1)=1.1\times10^{-13}\,\mathrm{s}$ or equivalently Dingle temperatures of $T_D(W1)=3.2\,\mathrm{K}$ and $T_D(H1)=11\,\mathrm{K}$ respectively \cite{Dingle52}. Taking into account the theoretical Fermi velocity of both pockets $v_F(W1)=296\times10^3\,\mathrm{m/s}$ and $v_F(W1)=755\times10^3\,\mathrm{m/s}$ we obtain a charge carrier mean-free path of $l=(95\pm15)\,\mathrm{nm}$. The corresponding dHvA electron and hole mobilities $\mu=\tau e/m^*$ are $\mu_\mathrm{e}=1.2\times10^4\,\mathrm{cm}^2\mathrm{/Vs}=1.2\,\mathrm{T}^{-1}$ and $\mu_\mathrm{e}=1.9\times10^4\,\mathrm{cm}^2\mathrm{/Vs}=1.9\,\mathrm{T}^{-1}$.

In summary, we observed quantum oscillations from three types of FS pockets in TaAs. By considering the angular dependence of all observed QO frequency branches, we were able to obtain a unique solution for the FS topology based on additional DFT calculations. The FS consists of three pairs of topological Weyl pockets and two trivial hole pockets aligned along each nodal ring of the tetragonal Brillouin zone. In TaAs, $E_\mathrm{F}$ is such that each Weyl point is surrounded by a distinct FS pocket. Compared to the other members of the non-centrosymmetric monopnictides \cite{Arnold15TaP,Klotz16}, TaAs is the first Weyl semimetal with confirmed chiral quasiparticles at $E_\mathrm{F}$. Thus TaAs is a prime candidate for the observation of chiral effects such as the Alder-Bell-Jackiw anomaly.

\section{Acknowledgements}

The authors would like to acknowledge the Max-Planck society for their support of the MPRG.


\end{document}